\documentclass[aps,pre,twocolumn,superscriptaddress,longbibliography,preprintnumbers,10pt]{revtex4-2} 

\usepackage{amsmath}
\usepackage{amsfonts}
\usepackage{amssymb}
\usepackage{graphicx}
\usepackage{color}
\usepackage{accents}
\usepackage[colorlinks=true, citecolor=cyan]{hyperref}
\usepackage{enumitem}
\usepackage{bbold}

\newcommand{\Tr}{\mbox{Tr}}

\newcommand{\ket}[1]{|#1\rangle}
\newcommand{\bra}[1]{\langle #1|}

\makeatletter 
\newsavebox{\@brx}
\newcommand{\llangle}[1][]{\savebox{\@brx}{\(\m@th{#1\langle}\)}%
	\mathopen{\copy\@brx\kern-0.5\wd\@brx\usebox{\@brx}}}
\newcommand{\rrangle}[1][]{\savebox{\@brx}{\(\m@th{#1\rangle}\)}%
	\mathclose{\copy\@brx\kern-0.5\wd\@brx\usebox{\@brx}}}
\makeatother

\newlength{\dhatheight} 

\newcommand{\qed}{\nobreak \ifvmode \relax \else
	\ifdim\lastskip<1.5em \hskip-\lastskip
	\hskip1.5em plus0em minus0.5em \fi \nobreak
	\vrule height0.75em width0.5em depth0.25em\fi}

\graphicspath{{./figures/}}

\begin{document}
	
	\title{Taking the temperature of quantum many-body scars}
	
	\author{Phillip C. Burke}
	\email[]{phillip.cussenburke@ucd.ie}
	\affiliation{School of Physics, University College Dublin, Belfield, Dublin 4, Ireland}
	\affiliation{Centre for Quantum Engineering, Science, and Technology, University College Dublin, Dublin 4, Ireland}
	
	\author{Shane Dooley}
	\email[]{dooleysh@stp.dias.ie}
	\affiliation{Dublin Institute for Advanced Studies, School of Theoretical Physics, 10 Burlington Rd, Dublin, Ireland}
	
	\date{\today}
	
	\begin{abstract}
		A quantum many-body scar is an eigenstate of a chaotic many-body Hamiltonian that exhibits two seemingly incongruous properties: its energy eigenvalue corresponds to a high temperature, yet its entanglement structure resembles that of low-temperature eigenstates, such as ground states. 
		Traditionally, a temperature is assigned to an energy \emph{eigenvalue} through the textbook canonical temperature-energy relationship. However, in this work, we use the \emph{eigenstate subsystem temperature} -- a recently developed quantity that assigns a temperature to an energy \emph{eigenstate}, based on the structure of its reduced density matrix.
		For a thermal state, the eigenstate subsystem temperature is approximately equal to its canonical temperature. 
		Given that quantum many-body scars have a ground-state-like entanglement structure, it is not immediately clear that their eigenstate subsystem temperature would be close to their canonical temperature. 
		Surprisingly, we find that this is the case: the quantum many-body scars have approximate ``knowledge'' of their position in the spectrum encoded within their state structure.
		

	\end{abstract}
	
	
	\maketitle
	
	
	
	
	
	
	
	One of the fundamental problems of quantum many-body physics is understanding how thermalization arises in closed systems -- and when it fails to occur. In general, quantum chaotic systems tend to thermalize, while integrable systems can fail to do so due to the presence of an extensive number of local conserved quantities. However, in recent years, a novel mechanism has been discovered that can prevent thermalization in chaotic systems: the presence of special Hamiltonian eigenstates known as quantum many-body scars (QMBS) \cite{Turner_Scars_Nat2018, SerbynAbaninPapic_QMBS_Nat2021, Cha-23b, Cho-19, Doo-20b, Mou-20a, Des-22a, Doo-22a, Mou-22c, Log-24a}. 
	
	QMBS are typically understood in the context of the eigenstate thermalization hypothesis (ETH), which posits a specific structure for the eigenstates $\hat{H}\ket{E} = E \ket{E}$ of thermalising systems \cite{Deutsch_PRA1991, Srednicki_PRE1994, Srednicki_ThermalFluc_JoPA1996, Srednicki_ThermalEquil_JoPA1999, Rigol_thermalization_Nat2008, RigolSrednicki_Thermalization_2012, Reimann_NJP2015, AlessioRigol_Chaos_AdvPhys2016, Deutsch_RepProgPhys2018, Mori_Ikeda_Ueda_thermalizationreview_JPB2018}. Broadly speaking, the ETH suggests that for a small (local) subsystem $S$, the reduced eigenstate density matrix $\hat{\rho}_S(E) = \Tr_{\bar{S}} (\ket{E}\bra{E})$ is approximately equal to the corresponding canonical density matrix restricted to $S$: \begin{equation} \hat{\rho}_S (E) \approx \hat{\sigma}_S (\beta_C), \label{eq:rough_ETH} \end{equation} where $\hat{\sigma}_S(\beta_C) = \Tr_{\bar{S}} [\hat{\sigma}(\beta_C)]$ and $\hat{\sigma}(\beta_C) = e^{-\beta_C \hat{H}} / \Tr [e^{-\beta_C \hat{H}}]$ is the canonical (or Gibbs) density matrix. If the relation in Eq.~\eqref{eq:rough_ETH} holds for a given eigenstate $\ket{E}$, then the eigenstate is considered \emph{thermal}. However, in certain models, there can exist a small number of nonthermal eigenstates, which are embedded in a background of thermal eigenstates at finite energy density. These nonthermal eigenstates are referred to as QMBS.
	
\begin{figure} 
		\centering
		\includegraphics[width=0.42\columnwidth]{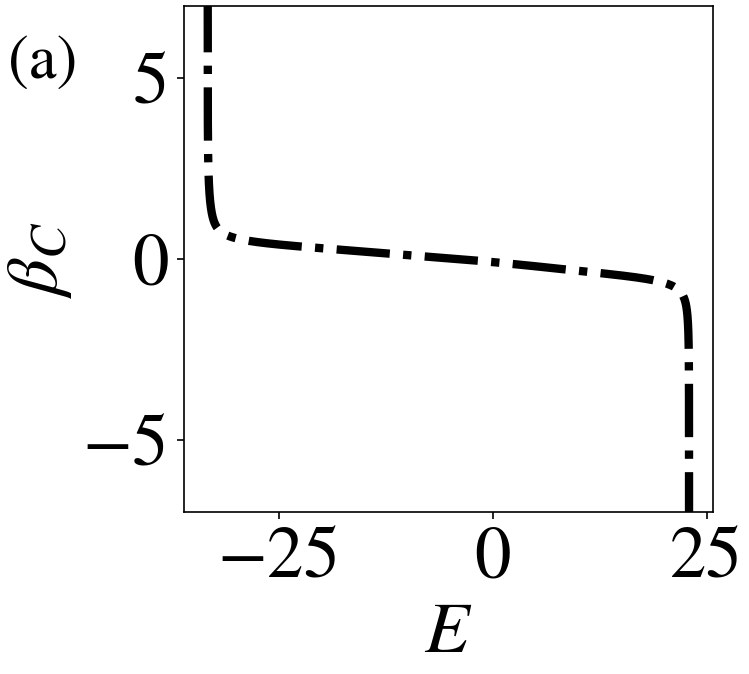}
		\includegraphics[width=0.56\columnwidth]{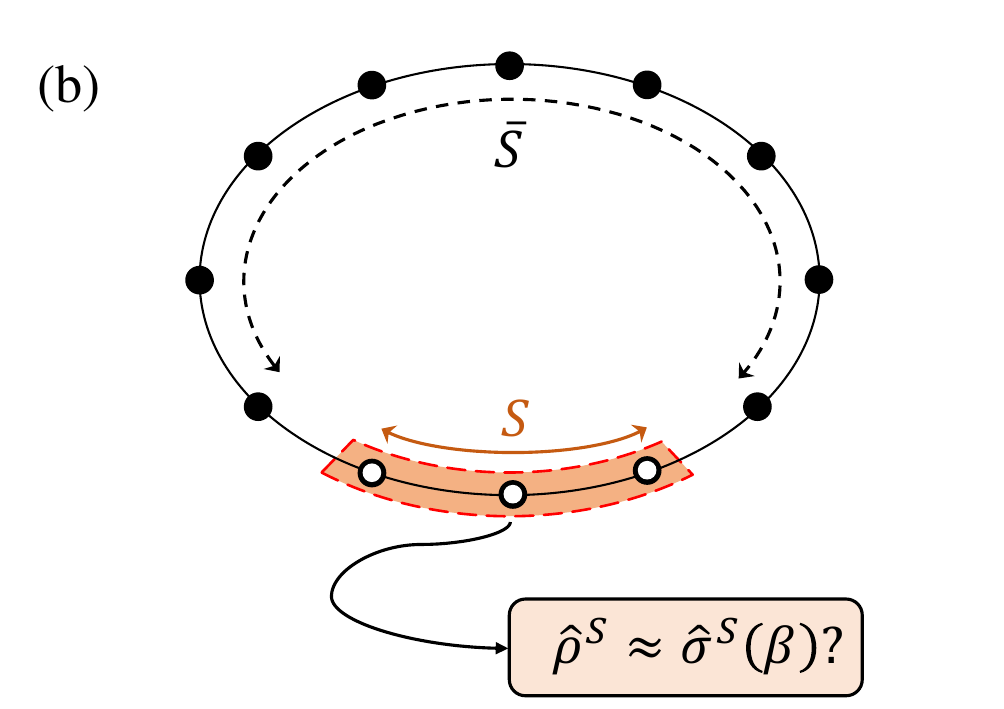}%
		\caption{(a) An example of a canonical temperature ($\beta_C)$ versus energy ($E$) curve (Eq.~\eqref{eq_canonical_beta}) for the `projected' uniform-field \textit{XXZ} spin-1/2 chain with $N=18$, within the $k=0$ momentum sector. (b) Illustration of a spin chain, divided into the subsystem $S$ and its complement $\bar{S}$.}
		\label{fig:illustrations}
	\end{figure}

	
	For the relationship in Eq.~\eqref{eq:rough_ETH} to hold, how does one know which energy eigenvalue $E$ to choose on the left-hand side, and which canonical temperature $\beta_C$ to choose on the right-hand side? The answer is that Eq.~\eqref{eq:rough_ETH} implicitly assumes a correspondence between $E$ and $\beta_C$ that is provided through the canonical temperature-energy relationship:
	\begin{equation}
		E \ = \ \Tr [\hat{\sigma}(\beta_C)\hat{H}] \ = \ \frac{\sum_{E'} e^{-\beta_C E'}E'}{\sum_{E'} e^{-\beta_C E'}} .
		\label{eq_canonical_beta}
	\end{equation}
	An example showing the correspondence between $\beta_C$ and $E$ for a particular Hamiltonian is shown in Fig.~\ref{fig:illustrations}(a). 

	However, the ETH, Eq.~\eqref{eq:rough_ETH}, represents a correspondence -- not only between the energy eigenvalue $E$ and the canonical temperature $\beta_C$ -- but also between the structure of the thermal eigenstate $\ket{E}$ and the canonical density matrix $\hat{\sigma}(\beta_C)$. This suggests that the canonical temperature $\beta_C$ is also encoded in the structure of the thermal eigenstate $\ket{E}$. 
	This idea was formalized in a recent work \cite{Bur-23a}, which assigned a temperature to energy eigenstates through the structure of their reduced density matrices, referred to as the \emph{eigenstate subsystem temperature} $\beta_S(\ket{E})$. It was demonstrated that the difference $\delta\beta(E) = \beta_S(\ket{E}) - \beta_C(E)$ for thermal eigenstates (i.e., those which follow the ETH) fluctuates around zero, with fluctuations vanishing in the thermodynamic limit for non-integrable systems, as one might expect from an ETH perspective. This showed quantitatively that a thermal eigenstate $\ket{E}$ and its energy eigenvalue $E$ independently encode the temperature but in a consistent manner.
	
	
	
	So, what is the eigenstate subsystem temperature encoded in a QMBS? In addition to their violation of the ETH, QMBS are characterized by their sub-volume law entanglement entropy and are typically well approximated by matrix product states (MPS) with low bond dimension \cite{Lin-19, Cha-20b, Mou-20a}. This is in stark contrast with thermal eigenstates, which follow a volume-law scaling of entanglement entropy, and cannot be described efficiently with MPS. In this sense, QMBS have a state structure that is much more similar to ground states of gapped local Hamiltonians, which are also well approximated by low bond dimension MPS \cite{Eis-10}, than to high-temperature thermal eigenstates. 
	
	
	An interesting question, therefore, is whether the eigenstate subsystem temperature $\beta_S(\ket{E})$ (which is determined by the eigenstate $\ket{E}$ rather than the eigenvalue $E$) can differentiate QMBS from the neighboring thermal eigenstates in the spectrum. At first glance, the two most extreme possibilities are: (1) The QMBS have an eigenstate subsystem temperature that is significantly different from their canonical temperature $\delta\beta(E) \equiv \beta_S(\ket{E}) - \beta_C(E) \not\approx 0$; this outcome might be expected considering the ground-state-like entanglement structure of the QMBS; or (2) The QMBS have an eigenstate subsystem temperature that is close to the canonical temperature $\delta\beta(E) = \beta_S(\ket{E}) - \beta_C(E) \approx 0$, despite their nonthermal structure.
	
	Surprisingly, for the models considered, we find an answer that lies somewhere between these two possibilities. On one hand, a QMBS state in a typical Hamiltonian (within our setting) has an eigenstate temperature that is \textit{close} to the canonical temperature. This is interesting because, despite the violation of Eq.~\eqref{eq:rough_ETH} and the large distance between the QMBS reduced density matrix and the reduced canonical density matrix, there appears to be some approximate ``knowledge'' of the canonical temperature encoded in a QMBS. On the other hand, $\delta\beta(E)$ differentiates QMBS from thermal eigenstates in the rate at which $\delta\beta(E)$ approaches zero, which is much more rapid for thermal eigenstates.
	
	
	
	\textit{Subsystem Eigenstate temperature--} It was shown recently in Ref.~\cite{Bur-23a} that it is possible to assign a temperature to the energy eigenstates $\ket{E}$ of a Hamiltonian $\hat{H}$ through the so-called \emph{subsystem temperature}. This is the temperature $\beta_S$ that minimizes the distance between the reduced canonical density matrix $\hat{\sigma}_{S} = \Tr_{\bar{S}} [\hat{\sigma}(\beta)]$ and the reduced eigenstate density matrix $\hat{\rho}_S = \Tr_{\bar{S}}\ket{E}\bra{E}$, where $S$ is a subsystem and $\bar{S}$ is the complement of the subsystem being traced over, i.e., 
	\begin{equation} 
		\beta_S(\ket{E}) \ = \ \text{argmin}_\beta [d_1( \hat{\rho}_{S}, \hat{\sigma}_{S}(\beta))] .
		\label{eq:subsystem_temp}
	\end{equation} 
	Here, $d_1(\hat{X}, \hat{Y}) = \|\hat{X}-\hat{Y}\|_1$ is the trace distance and $\|\hat{X}\|_1 = \Tr\sqrt{\hat{X}^\dagger\hat{X}}$ is the trace norm (Schatten 1-norm).
	It is important to note that the reduced canonical density matrix $\hat{\sigma}_{S}$ defined here is, in general, not equal to the canonical density matrix $e^{-\beta \hat{H}_S}/\Tr e^{-\beta \hat{H}_S}$ constructed using the Hamiltonian $\hat{H}_S$ restricted to the subsystem $S$ \cite{Gelin_Thoss_subensemble_PRE2009, Hilt_HMF_PRE2011, GallegoEisert_BeyondWeakCoup_NJoP2014, BurkeNakerstHaque_HMF_PRE2024}.
	
	
	In general, the subsystem temperature $\beta_S$ for an eigenstate $\ket{E}$ is not equal to the canonical temperature $\beta_C$ for the corresponding eigenvalue $E$, i.e., $\delta\beta(E) = \beta_S(\ket{E}) - \beta_C(E) \neq 0$. However, for a thermal eigenstate (i.e., an eigenstate that obeys the ETH Eq.~\eqref{eq:rough_ETH}) we expect $\hat{\rho}_S (E) \approx \hat{\sigma}_S (\beta_C)$ and $\beta_S \approx \beta_C$. Moreover, for thermal states we expect the difference $\delta\beta(E) = \beta_S(\ket{E}) - \beta_C(E)$ to approach zero in the thermodynamic limit \cite{Bur-23a}. For non-thermal states, e.g., QMBS, we have $\Tr_{\bar{S}} (\ket{E}\bra{E}) \not\approx {\hat{\sigma}_S}(\beta_C)$ and so we no longer necessarily expect $\beta_S \approx \beta_C$.
	
	
	\textit{Model --} 
	We construct models with QMBS through the projector-embedding approach of Shiraishi and Mori \cite{Shi-17}. Specifically, we consider a chain of $N$ spin-$1/2$ particles with a local Hamiltonian of the form: \begin{equation} \hat{H} = \sum_{n=0}^{N-1} \hat{P}_{n,n+1} \hat{h}_{n,n+1} \hat{P}_{n,n+1} . \label{eq:H} \end{equation} 
	Here, $\hat{h}_{n,n+1} = \hat{\mathbb{1}}^{\otimes n} \otimes \hat{h} \otimes \hat{\mathbb{1}}^{\otimes (N - n - 2)}$ where $\hat{h}$ is a two-spin Hermitian matrix acting on the neighboring spins labeled $n$ and $n+1$, and $\hat{\mathbb{1}}$ is the spin-1/2 identity operator, which acts on all other spins. Similarly, $\hat{P}_{n,n+1} = \hat{\mathbb{1}}^{\otimes n} \otimes \hat{P} \otimes \hat{\mathbb{1}}^{\otimes (N - n - 2)}$ where $\hat{P}^2 = \hat{P}$ is a two-spin projector acting on spins $n$ and $n+1$. Throughout this work, we choose the two-spin projector $\hat{P} = \hat{\mathbb{1}} \otimes \hat{\mathbb{1}} - \ket{0}\bra{0} \otimes \ket{0}\bra{0}$. This ensures that the product state $\ket{E_{\rm QMBS}} = \ket{0}^{\otimes N}$ is annihilated by the Hamiltonian for any possible choice of $\hat{h}$, and is therefore a zero-energy eigenstate $\hat{H} \ket{E_{\rm QMBS}} = 0$. We note that, although the eigenstate $\ket{E_{\rm QMBS}} = \ket{0}^{\otimes N}$ is always fixed at zero energy $E_{\rm QMBS} = 0$, its relative position in the spectrum varies depending on the choice of $\hat{h}$. However, if $E_{\rm QMBS}=0$ is far from the edges of the spectrum, and moreover if $\hat{H}$ is non-integrable, then $\ket{E_{\rm QMBS}} = \ket{0}^{\otimes N}$ is regarded as a QMBS. 
	
	The Hamiltonian in Eq.~\eqref{eq:H} has periodic boundary conditions and is translation invariant, i.e., $[\hat{T},\hat{H}] = 0$, where $\hat{T}$ is the one-site translation operator \footnote{There have been various numerical and analytical confirmations of the ETH within translation-invariant systems \cite{SugimotoErdos_ETH_TIS_JoSP2023, Huang_SubsystemETH_PRE2024}}. The Hamiltonian eigenstates $\hat{H}\ket{E_\alpha} = E_\alpha \ket{E_\alpha}$ are therefore simultaneous eigenstates of the translation operator $\hat{T}\ket{E_\alpha} = e^{i2 \pi k /N} \ket{E_\alpha}$, and can be labeled according to their quasi-momentum $k \in \mathbb{Z}$. Our QMBS $\ket{E_{\rm QMBS}} = \ket{0}^{\otimes N}$ is itself translation invariant, and so falls in the $k=0$ symmetry sector of the Hamiltonian. Throughout this work, we choose the subsystem $S$ (e.g., for computing the reduced density matrix $\hat{\rho}_S(E) = \Tr_{\bar{S}} \ket{E}\bra{E}$) to be two neighbouring spins in the chain.
	
\begin{figure} 
		\includegraphics[width=\columnwidth]{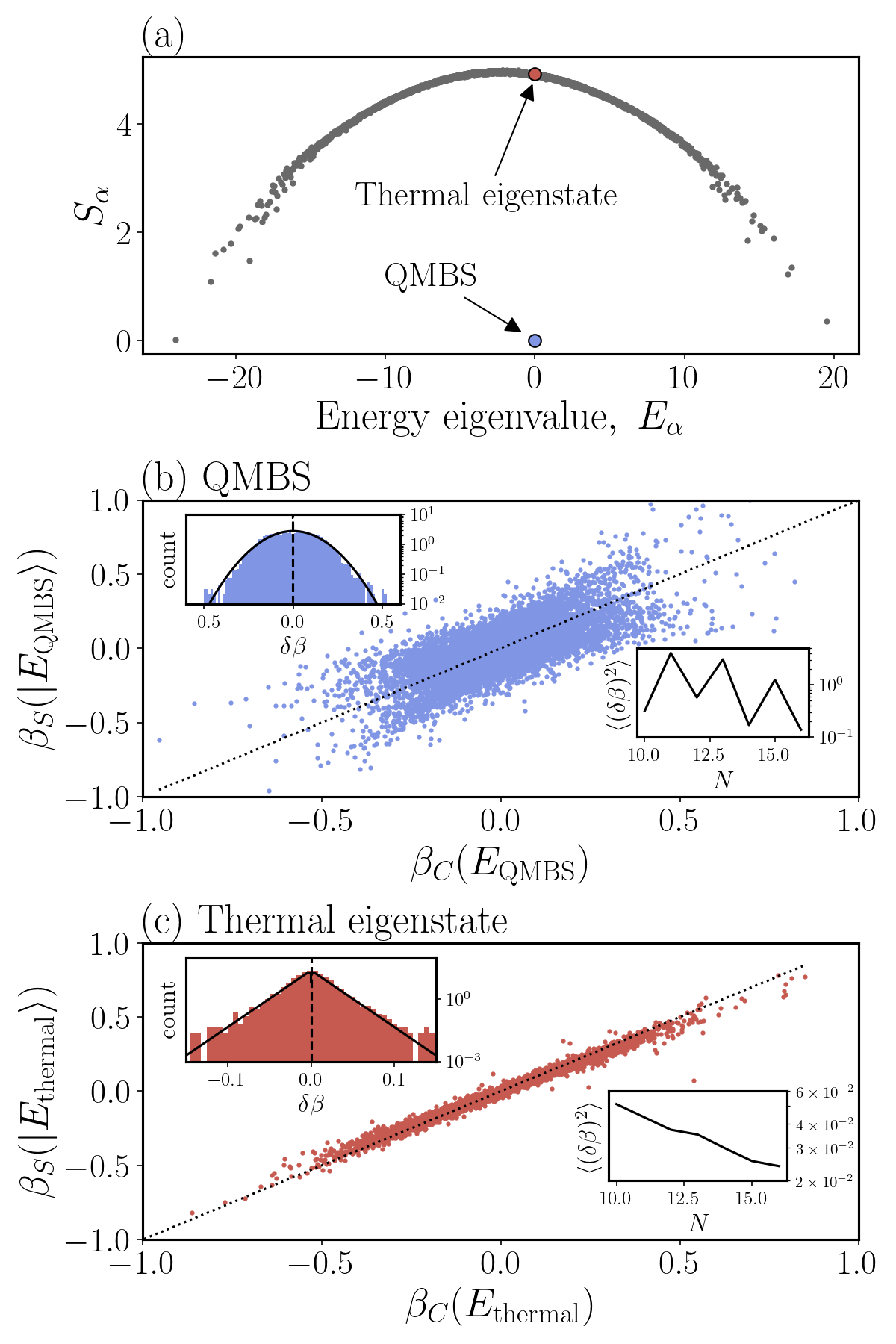}%
		\caption{
			We generate the Hamiltonian $\hat{H}$ (Eq.~\eqref{eq:H}) of a chain of $N=16$ spin-1/2 particles, choosing the local two-spin Hermitian matrix $\hat{h}$ at random from the Gaussian unitary ensemble (GUE). (a) For a typical instance of the Hamiltonian $\hat{H}$, the entanglement entropy of each eigenstate is plotted against its corresponding eigenvalue. (b) Each data point $(\beta_S (\ket{E_{\rm QMBS}})$, $\beta_C (E_{\rm QMBS}))$ is calculated for the QMBS $\ket{E_{\rm QMBS}}$ of a randomly generated Hamiltonian $\hat{H}$. (c) Each data point $(\beta_S (\ket{E_{\rm thermal}}), \beta_C (E_{\rm thermal}))$ is calculated for the representative thermal eigenstate $\ket{E_{\rm thermal}}$ of a randomly generated Hamiltonian $\hat{H}$. The upper insets show the histograms of $\delta\beta = \beta_S - \beta_C$. The dashed vertical lines show that the average $\langle \delta\beta \rangle \approx 0$ is close to zero in both cases. For thermal states, the distribution of $\delta\beta$ decays exponentially around $\delta\beta \approx 0$, while for QMBS it decays as a Gaussian (solid black lines). 
			The lower-right insets show the statistical variance $\langle (\delta \beta)^2\rangle$ versus system size $N$.
		}
		\label{fig:delta_beta_for_random_models}
	\end{figure}

	To explore a variety of generic models with the QMBS $\ket{E_{\rm QMBS}} = \ket{0}^{\otimes N}$, we generate the two-spin Hamiltonian term $\hat{h}$ at random from the Gaussian unitary ensemble (GUE) of two-spin Hermitian matrices. This will typically result in a non-integrable Hamiltonian $\hat{H}$. An example for a particular randomly generated instance of $\hat{h}$ is shown in Fig.~\ref{fig:delta_beta_for_random_models}(a), where we plot the entanglement entropy $S_\alpha$ for every eigenstate $\ket{E_\alpha}$ of the corresponding Hamiltonian $\hat{H}$. The entanglement entropy is defined as $S_\alpha = -\Tr [\hat{\rho}^{(\alpha)}\log \hat{\rho}^{(\alpha)}]$, where $\hat{\rho}^{(\alpha)} = \Tr_{N/2} \ket{E_\alpha} \bra{E_\alpha}$ is the reduced density matrix of the eigenstate on one contiguous half of the spin chain. The QMBS is clearly visible as an outlier with $S_\alpha = 0$ (blue marker), while all other eigenstates away from the edges of the spectrum are thermal and have a high entanglement entropy. 
	
	
	In certain cases, it will also be useful to fix a particular $\hat{h}$. We choose the spin-1/2 XXZ interaction with an additional (uniform) transverse magnetic field: 
	\begin{equation} 
		\hat{h} = b (\hat{s}^x \otimes \hat{\mathbb{1}} + \hat{\mathbb{1}} \otimes \hat{s}^x ) +  J (\hat{s}^x \otimes \hat{s}^x + \hat{s}^y \otimes \hat{s}^y ) + \Delta \hat{s}^z \otimes \hat{s}^z, 
		\label{eq:xxz_ham}
	\end{equation} 
	where $\hat{s}^\mu$, $\mu \in \{ x,y,z \}$ are the standard spin-1/2 Pauli operators.
	
	Our goal is to explore the deviation between the subsystem temperature and the canonical temperature $\delta\beta (E_{\rm QMBS}) = \beta_S (\ket{E_{\rm QMBS}}) - \beta_C (E_{\rm QMBS})$ for the QMBS. However, as a reference, we also compute the deviation $\delta\beta (E_{\rm thermal})$ for a representative thermal state $\ket{E_{\rm thermal}}$, which we always choose to be the eigenstate of $\hat{H}$ with the smallest positive energy, i.e., $\ket{E_{\rm thermal}}$ is the neighbouring thermal eigenstate to $\ket{E_{\rm QMBS}}$ in terms of energy [see Fig.~\ref{fig:delta_beta_for_random_models}(a)].

	\textit{Results --} 
	For each randomly generated instance of the Hamiltonian $\hat{H}$, we numerically compute its energy eigenvalues and eigenstates $\hat{H}\ket{E_\alpha} = E_\alpha \ket{E_\alpha}$ by exact diagonalization. If the energy level spacing statistics align with those expected for a translational invariant quantum chaotic system \cite{AtasBogomolny_LevelStats_PRL2013}, and if the QMBS energy $E_{\rm QMBS} = 0$ is in the central half of the spectrum, we proceed to compute the QMBS subsystem temperature $\beta_S (\ket{E_{\rm QMBS}})$, its canonical temperature $\beta_C (E_{\rm QMBS})$, and the difference $\delta\beta (E_{\rm QMBS}) = \beta_S (\ket{E_{\rm QMBS}}) - \beta_C (E_{\rm QMBS})$. 
	
	The results for $\sim 7,000$ randomly generated $\hat{h}$ are plotted in Fig.~\ref{fig:delta_beta_for_random_models}(b), with the upper-left inset showing the histogram of generated $\delta\beta (E_{\rm QMBS})$ values. 
	Immediately, we can see that the QMBS canonical temperature $\beta_C (E_{\rm QMBS})$ and its subsystem temperature $\beta_S (\ket{E_{\rm QMBS}})$ are correlated. In other words, despite the QMBS having a similar entanglement structure to ground states (blue marker in panel (a)) and, on average, a large distance to $\hat{\sigma}_S$ (see Supplemental Material, Ref. \cite{SuppMat}), if the QMBS canonical temperature is high, then its subsystem temperature is also high. The histogram in the inset to Fig.~\ref{fig:delta_beta_for_random_models}(b) shows that, if we take the average over randomly generated models, we have $\langle \delta\beta \rangle \approx 0$ (the dashed vertical black line). The inset also shows that, near $\delta\beta \approx 0$, the differences $\delta\beta$ are normally distributed $P(\delta\beta) \sim \exp [-\frac{1}{2}(\delta\beta)^2 / \langle (\delta\beta)^2\rangle ]$ (the solid black) where $\langle (\delta\beta)^2 \rangle$ is the statistical variance of the distribution of $\delta\beta$. The lower-right inset of Fig.~\ref{fig:delta_beta_for_random_models}(b) shows the dependence of the statistical variance $\langle (\delta\beta)^2 \rangle$ on the system size $N$. There is a clear oscillation in $\langle (\delta\beta)^2 \rangle$, although, focusing on either odd $N$ or even $N$, the statistical variance appears to decrease slowly with system size, indicating that the correlation between $\beta_S$ and $\beta_C$ becomes tighter as system size increases. However, this evidence is not conclusive, and further numerical simulations for larger system sizes would be needed to verify this. (In Ref. \cite{SuppMat} we also show the scaling with $N$ of several dimensionless correlation measures, including the Pearson correlation coefficient.)
	
	We compute the same quantities for the representative thermal state $\ket{E_{\rm thermal}}$ of each random Hamiltonian, with the results shown in Fig.~\ref{fig:delta_beta_for_random_models}(c). Consistent with the findings of Ref.~\cite{Bur-23a}, these thermal states also have a subsystem temperature that is close to the canonical temperature $\beta_C$. However, compared to the corresponding data for QMBSs, we observe a stronger correlation between $\beta_S$ and $\beta_C$ for thermal states. 
	
	The upper-left inset of Fig.~\ref{fig:delta_beta_for_random_models}(c) shows the distribution of $\delta\beta$ as a histogram. Notably, the distribution exhibits an exponential decay near $\delta\beta=0$ (following $P(\delta\beta) \sim \exp [-\lambda \delta\beta / \sqrt{\langle (\delta\beta)^2\rangle}]$ with a fitting parameter $\lambda$), in contrast to the normal distribution for the QMBS in Fig.~\ref{fig:delta_beta_for_random_models}(b). The statistical variance for the thermal state shows a clear decrease as the system size increases, as shown in the lower-right inset of Fig.~\ref{fig:delta_beta_for_random_models}(c)
	
	\begin{figure}[htb]
		\centering 
		\includegraphics[width=\columnwidth]{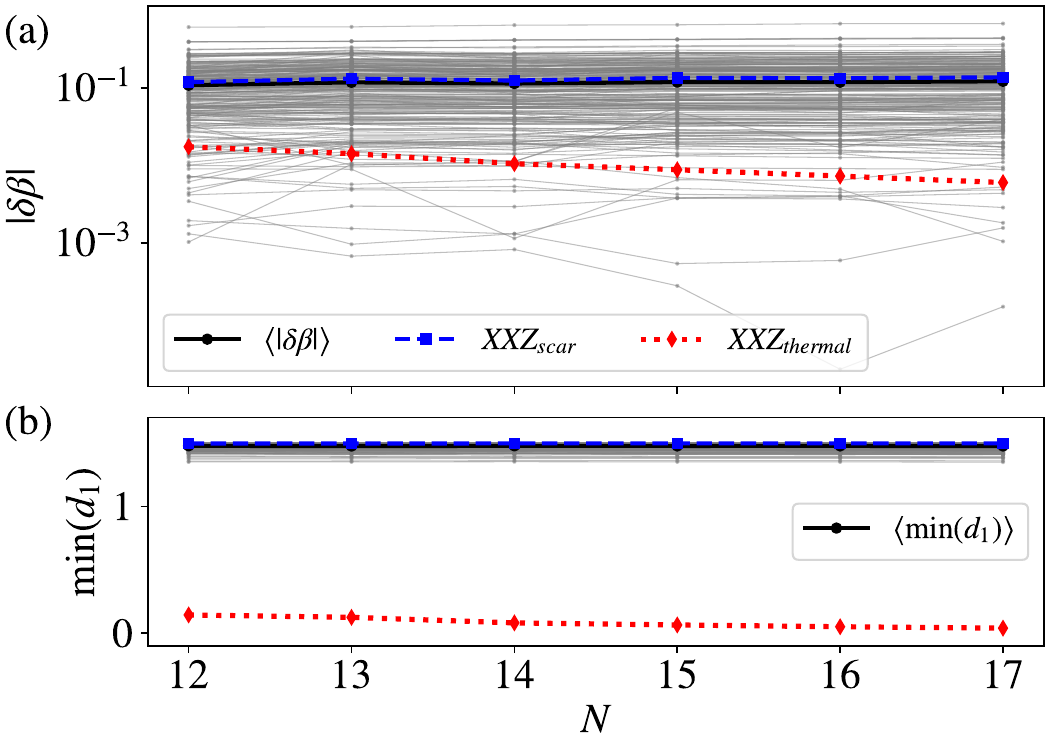}%
		\caption{(a) $\delta\beta$ versus $N$, (b) $\min(d_1)$ versus $N$ -- Grey lines indicate individual random instances of the random local model, and the solid black line denotes the average over the $\sim300$ realizations. The dashed blue and dotted red lines correspond to a QMBS state and an average over highly excited thermal states of the uniform field \textit{XXZ} chain respectively.}
		\label{fig:deltabeta}
	\end{figure}
	
	
	Thus far, we have computed statistical quantities from data generated across many realizations of randomly generated models. This allowed us to infer the typical behavior of $\delta\beta$ for a typical chaotic Hamiltonian at a given system size.  
	It is also worthwhile to investigate how $|\delta\beta|$ changes as the system size $N$ increases for a \emph{fixed} $\hat{h}$. The blue line in Fig.~\ref{fig:deltabeta}(a) shows this for the QMBS in a model with $\hat{h}$ chosen to be the transverse field XXZ interaction (Eq.~\eqref{eq:xxz_ham}), while each grey line shows it for $\hat{h}$ chosen at random (the black line shows the average of all the grey lines). In each case, for the QMBS we do not observe decay in $|\delta\beta|$ as $N$ increases, up to the largest system size accessible to us numerically. For comparison, we show $|\delta\beta|$ averaged over highly excited thermal (non-QMBS) states in the same system. In this case, $|\delta\beta|$ decays with increasing system size, consistent with expectations from the ETH. 
	
	
	In Fig.~\ref{fig:deltabeta}(b), we also plot $\min(d_1)$ corresponding to $\beta_S$, providing a measure of the state's degree of thermality. As expected, a clear distinction emerges between QMBS states (both in the random model and the projected XXZ chain) and the highly excited thermal states of the projected XXZ chain. 
	Although the thermal states become increasingly thermal with growing $N$, the QMBSs remain significantly distant from any thermal state. In particular, the distance measure for the QMBS remains approximately constant as $N$ increases, suggesting their robustness in the thermodynamic limit.
	
	\textit{Summary \& Outlook --} 
	In this work, we have demonstrated that -- despite the nonthermal entanglement structure of QMBS embedded in the high-energy spectrum of a chaotic system -- these eigenstates retain in their reduced density matrices an approximate knowledge of their position in the spectrum.
	Specifically, we have shown that the subsystem temperature, $\beta_S$, which characterizes the closest (local) thermal states to the QMBS, remains correlated with the corresponding canonical temperature, $\beta_C$. This occurs despite QMBS exhibiting a large distance from any (local) thermal state, quantified by the distance measure $d_1$.
	
	
	Although we consider a large variety of models with QMBS, we only consider a single QMBS of the form $\ket{E_{\rm QMBS}} = \ket{0}^{\otimes N}$. Future research could extend our results to QMBS with different entanglement structures, such as those described by matrix product states (MPS) \cite{Mou-20a}, QMBS in the PXP model \cite{Turner_Scars_Nat2018}, or examine cases where multiple QMBS states coexist within the spectrum. Additionally, investigating whether similar eigenstate subsystem temperature properties emerge in other nonthermal states or regimes, e.g., many-body localized \cite{PalHuse_MBL_PRB2010, SierantVidmar_MBL_RepInPhys2025} or integrable systems \cite{CassidyRigol_GenTherm_PRL2011, AlessioRigol_Chaos_AdvPhys2016, Vidmar_GGE_IOP2016}, could provide further insight into the breakdown of thermalization.
	
	Finally, we note that in addition to the canonical temperature definition in Eq.~\eqref{eq_canonical_beta}, the standard relationship between temperature and thermodynamic entropy from statistical mechanics has also been used to define temperature in finite systems. The precise definition of entropy in this context remains an active area of interest \cite{Gurarie_Entropy_AJoP2007, TalknerHanggiMorillo_MicrocanFluctuation_PRE2008, DunkelHilbert_NegativeT_Nat2014, HilbertHanggi_ThermoIsolated_PRE2014, Campisi_MicroEntropy_PRE2015, Hanggi_Temperature_RoyalSoc2016, BurkeHaque_Entropy_PRE2023, GovindRajan_Entropy_JoPCL2024}. 
	For such a definition of temperature based on thermodynamic entropy, the presence of QMBS will not significantly alter the temperature. However, if one instead considers entropy measures such as entanglement entropy, the nonthermal structure of QMBS will have a profound effect on the temperature.

	\textit{Acknowledgments --}
	P.C.B acknowledges funding from Science Foundation Ireland through grant 21/RP-2TF/10019. S. D. acknowledges support through the SFI-IRC Pathway Grant 22/PATH-S/10812. The authors also wish to acknowledge the Irish Centre for High-End Computing (ICHEC) for the provision of computational facilities and support. 
	
	%

	
	\newpage
	
	\setcounter{page}{1} \renewcommand{\thepage}{S\arabic{page}}
	
	\setcounter{figure}{0}
	\renewcommand{\thefigure}{S\arabic{figure}}
	
	\setcounter{equation}{0}
	\renewcommand{\theequation}{S.\arabic{equation}}
	
	\setcounter{table}{0}
	\renewcommand{\thetable}{S.\arabic{table}}
	
	\setcounter{section}{0}
	\renewcommand{\thesection}{S.\Roman{section}}
	
	\renewcommand{\thesubsection}{S.\Roman{section}.\Alph{subsection}}

	\makeatletter
	\renewcommand*{\p@subsection}{}
	\makeatother

	\renewcommand{\thesubsubsection}{S.\Roman{section}.\Alph{subsection}-\arabic{subsubsection}}
	
	\makeatletter
	\renewcommand*{\p@subsubsection}{}  
	\makeatother
	
	\begin{center}
		\large{
			Supplemental Material for \\ \textit{Taking the temperature of quantum many-body scars}
		}
	\end{center}

	
	\section*{Overview}
	
	\begin{itemize}
		\item In Section \ref{supp_corr}, we present two alternative ways of studying the correlation between $\beta_S$ and $\beta_C$.
		\item In Section \ref{supp_dist}, we show the distance measure values corresponding to the $\beta_S$ data in the main text.
	\end{itemize}
	
	
	\section{Correlation between canonical temperature and subsystem temperature}\label{supp_corr}
	
	In Fig.~\ref{fig:delta_beta_for_random_models}(b) of the main text, we see a clear correlation between the canonical temperature $\beta_C$ and the subsystem temperature $\beta_S$ for the QMBS states. Similarly, in Fig.~\ref{fig:delta_beta_for_random_models}(c) we see a correlation between the canonical temperature and the subsystem temperature for a thermal state. In the insets of both figures, the tightness of the correlation is quantified by $\langle (\delta\beta)^2 \rangle$, the statistical variance of the distribution of $\delta\beta = \beta_S - \beta_C$. However, the spread $\langle (\delta\beta)^2 \rangle$ of the points may be difficult to interpret without some temperature scale for comparison (for example, does $\delta\beta = 0.2$ correspond to a large spread or a small spread?). In this section, we provide two alternative ways of representing the correlation, which are normalised to avoid this difficulty: the ``fraction-of-the-spectrum'' and the Pearson correlation coefficient.
	
	\subsection{Fraction-of-the-spectrum}
	
	
	As an alternative, and possibly more intuitive representation of the data in Fig.~\ref{fig:delta_beta_for_random_models}(b,c), we introduce the dimensionless ``fraction-of-the-spectrum'' quantities. First, we define the canonical energy fraction-of-the-spectrum as: \begin{equation} e_C (E) = \frac{E - E_{\rm min}}{E_{\rm max} - E_{\rm min}} , \end{equation} where $E_{\rm max}$ and $E_{\rm min}$ are the maximum and minimum energy eigenvalues of the Hamiltonian $\hat{H}$, respectively. It gives the fraction of the spectrum at which the energy $E$ appears, and can take values between $e_C(E_{\rm min}) = 0$ and $e_C(E_{\rm max}) = 1$. The canonical energy fraction-of-the-spectrum $e_S (E)$ transforms the canonical temperature $\beta_C (E)$ to a dimensionless quantity.
	
	Similarly we define the subsystem energy fraction-of-the-spectrum as: \begin{equation} e_S (\ket{E}) = \frac{\mathcal{E}(\ket{E}) - E_{\rm min}}{E_{\rm max} - E_{\rm min}} . \label{eq:e_S} \end{equation} The quantity $\mathcal{E}(\ket{E})$ is an energy associated with the subsystem temperature $\beta_S(\ket{E})$ of the eigenstate $\ket{E}$. It is defined as $\mathcal{E} = \Tr [\hat{\sigma}(\beta_S) \hat{H}]$. Finally, the subsystem energy $\mathcal{E}$ is translated into a fraction-of-the-spectrum via Eq.~\eqref{eq:e_S}.
	
	\begin{figure}[htb]
		\includegraphics[width=\columnwidth]{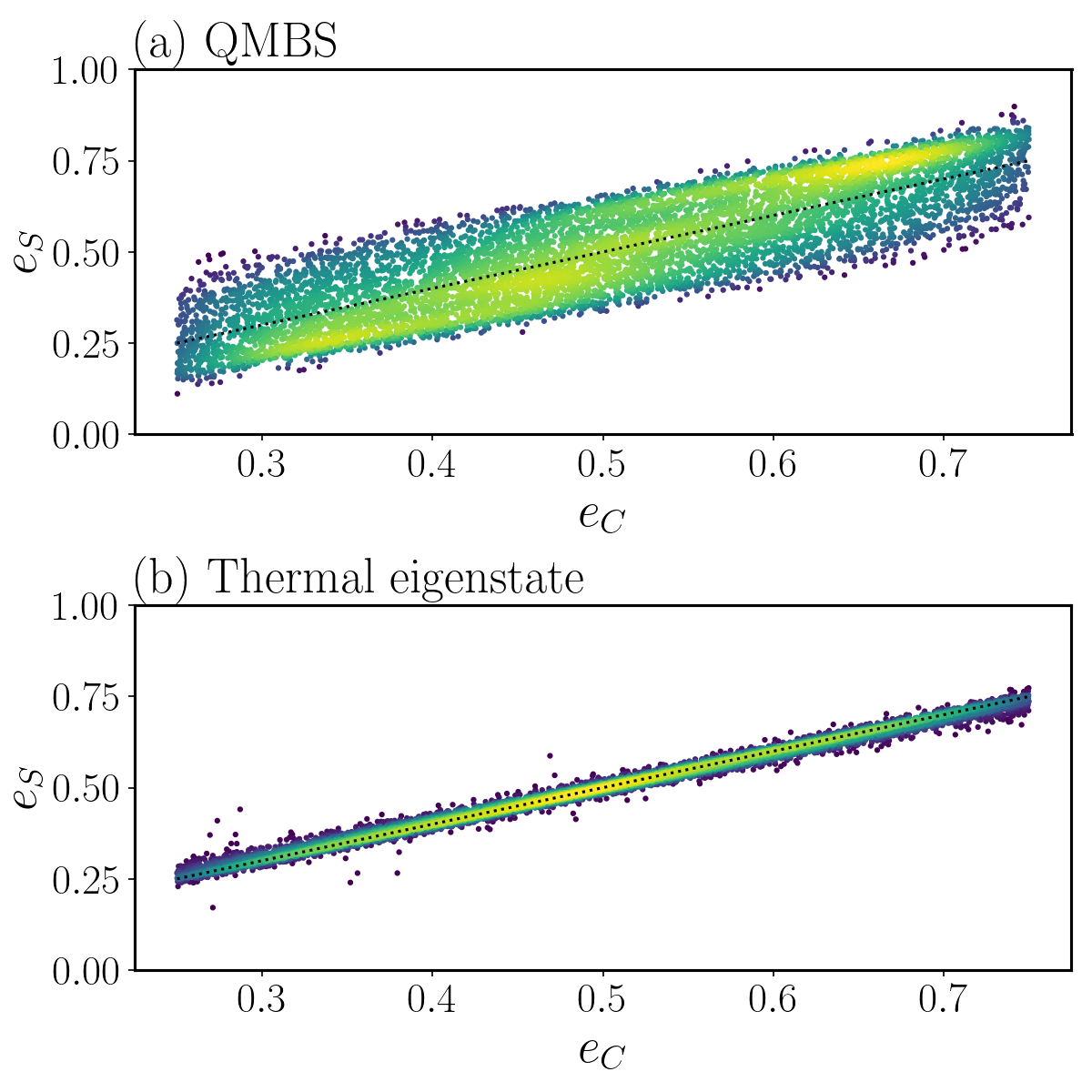}%
		\caption{The same data as in Fig.~\ref{fig:delta_beta_for_random_models}(b,c), translated into the ``fraction of the spectrum'': $e_S$ replacing $\beta_S$ and $e_C$ replacing $\beta_C$. The marker color indicates the density of surrounding data points: yellow for more density of data points and blue for lower density.
		}
		\label{fig:delta_e_for_random_models}
	\end{figure}
	
	The quantities $e_S$ and $e_C$ are plotted in Fig.~\ref{fig:delta_e_for_random_models}. We again see the correlation between the canonical energy and the subsystem energy, for both QMBS and thermal eigenstates, with a tighter correlation for thermal states. However, in this fraction-of-the-spectrum representation, we can also see clearly that mid-spectrum values of $e_C$ generally correspond to mid-spectrum values of $e_S$.
	
	\subsection{Pearson correlation coefficient}
	
	An alternative measure of the correlation between the canonical temperature $\beta_C$ and the subsystem temperature $\beta_S$ is the Pearson correlation coefficient: 
	\begin{equation} 
		\mathrm{corr} = \frac{\mathrm{cov}(\beta_C, \beta_S)}{\mathrm{std}(\beta_C) \mathrm{std}(\beta_s)} ,
	\end{equation} 
	where $\mathrm{cov}(\beta_C, \beta_S) = \langle (\beta_C - \langle\beta_C\rangle)(\beta_S - \langle\beta_S\rangle) \rangle $ is the covariance between $\beta_C$ and $\beta_S$, while $\mathrm{std}(\beta_s)$ and $\mathrm{std}(\beta_s)$ are the standard deviations of $\beta_C$ and $\beta_S$, respectively.
	The normalization by the standard deviations in the denominator results in a dimensionless quantity. It takes values in the range $-1 \leq \mathrm{corr} \leq 1$, with $\mathrm{corr} = 1$ corresponding to perfect correlation, $\mathrm{corr} = -1$ to perfect anti-correlation, and $\mathrm{corr} = 0$ to no correlation.
	
	\begin{figure}[htb]
		\includegraphics[width=\columnwidth]{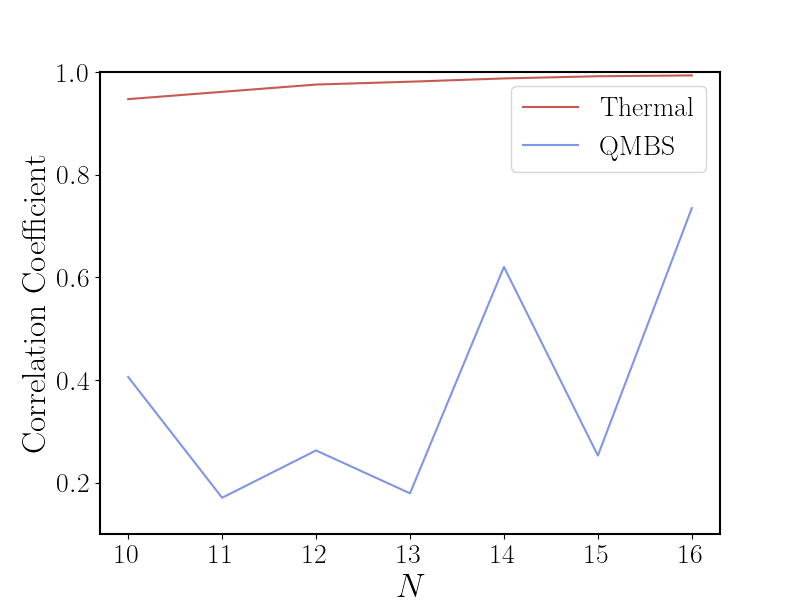}%
		\caption{The Pearson correlation coefficient for the canonical temperature $\beta_C$ and the subsystem temperature $\beta_S$, as a function of system size $N$. The $N=16$ points here correspond to the data in Figs.~\ref{fig:delta_beta_for_random_models}(b,c).
		}
		\label{fig:corr}
	\end{figure}
	
	In Fig.~\ref{fig:corr} we plot the correlation coefficient $\mathrm{corr}$ as a function of the system size $N$, for both the QMBS and the thermal state. We see a very strong correlation between $\beta_C$ and $\beta_S$ for the thermal state, which is expected considering that this is clearly visible in the data for $N=16$ presented in Fig.~\ref{fig:delta_e_for_random_models}(b). For the QMBS, there is an apparent oscillation in $\textsf{corr}$, although, focusing on either odd or even $N$, the correlation appears to increase slowly with system size (above $N=10$). This complements the results obtained in the lower-right inset to Fig.~\ref{fig:delta_beta_for_random_models}(b). However, the evidence is not conclusive, and further numerical simulations for larger system sizes would be needed to verify that the correlation continues to increase with system size.

	
	\begin{figure}[htb]
		\includegraphics[width=\columnwidth]{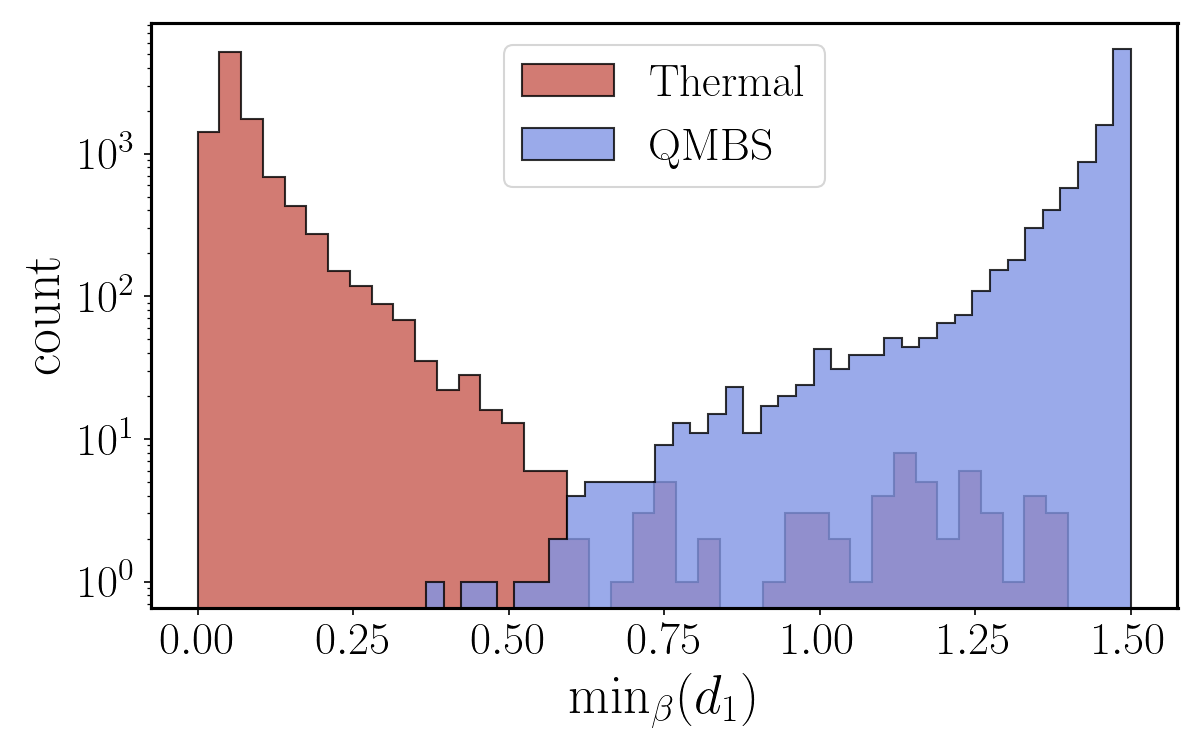}%
		\caption{Histograms of the minimum distance measures corresponding to the subsystem temperature data in Fig. \ref{fig:delta_beta_for_random_models} of the main text. 
		}
		\label{fig:distances_histogram}
	\end{figure}
	
	\section{Corresponding distance measure to subsystem temperature}\label{supp_dist}
	
	In the main text, we presented subsystem temperature data for many random realizations of our projected QMBS model. Here, we present the corresponding minimum distance value in Fig.~\ref{fig:distances_histogram}.
	We see that the histogram of the distances for thermal states is in stark contrast to that of the QMBSs. 
	For the thermal states, the mean value is close to zero with a skewed tail decaying towards positive values, while for QMBSs, the mean is close to $\sim1.5$ with a tail decaying towards zero.
	

	
	
\end{document}